\documentclass[useAMS,usenatbib]{mn2e}
\usepackage{graphicx,amssymb,color}
\usepackage[normalem]{ulem}

\title[Full-lifetime non-Kozai inclinations]
{Effects of non-Kozai mutual inclinations on two-planet system stability through all phases of stellar evolution}

\author[Veras, Georgakarakos, G\"{a}nsicke \& Dobbs-Dixon]{
Dimitri Veras$^{1,2}$\thanks{E-mail: d.veras@warwick.ac.uk}\thanks{STFC Ernest Rutherford Fellow},
Nikolaos Georgakarakos$^{3}$,
Boris T. G\"{a}nsicke$^{1,2}$,
Ian Dobbs-Dixon$^{3}$
\\
$^{1}$Centre for Exoplanets and Habitability, University of Warwick, Coventry CV4 7AL, UK
\\
$^{2}$Department of Physics, University of Warwick, Coventry CV4 7AL, UK
\\
$^{3}$New York University Abu Dhabi, Saadiyat Island, P.O. Box 129188, Abu Dhabi, UAE
}

\pubyear{2018}

\begin{document}
\label{firstpage}
\pagerange{\pageref{firstpage}--\pageref{lastpage}}
\maketitle

\begin{abstract}
Previous full-lifetime simulations of single-star multi-planet systems across all phases of stellar evolution have predominately assumed coplanar or nearly-coplanar orbits. Here we assess the consequences of this assumption by removing it and exploring the effect of giant branch mass loss on the stability of two-planet systems with small to moderate non-Kozai ($<40$ degrees) relative inclinations. We run nearly $10^4$ simulations over 14 Gyr for F-star, A-star and B-star planet hosts, incorporating main sequence stellar masses of 1.5, 2.0, 2.5, 3.0 and 5.0 solar masses, and initial planetary semimajor axis ratios that straddle their three-dimensional Hill stability limits. We find that the near-coplanar assumption can approximate well the stability frequencies and critical separations found for higher inclinations, except around strong mean-motion commensurabilities. Late instabilities -- after the star has become a white dwarf -- occur throughout the explored mutual inclination range. Consequently, non-Kozai mutual inclination should not be used as a predictive orbital proxy for determining which white dwarf multi-planet systems discovered by {\it Gaia} should represent high-priority follow-up targets for the detection of metal pollution and planetary debris discs.
\end{abstract}


\section{Introduction}

One way that the {\it Gaia} mission could transform the field of exoplanetary science
is by allowing us to measure the mutual inclinations of 
the orbits of an ensemble of planets residing a few au from their parent
stars \citep{peretal2014}. For decades, this elusive orbital 
parameter has inhibited our efforts to constrain 
dynamics \citep{butetal1999}, and has represented a frustrating
extra degree of freedom in multi-body modelling efforts.

Fortunately, {\it Gaia}'s exoplanet detections and inclination measurements will not be limited to main 
sequence host stars. The mission will likely also break new ground by discovering 
exoplanets orbiting white dwarfs at a distance of a few au. Our current wealth of knowledge of 
white dwarf planetary systems instead arises from the immediate circumstellar environment, within about
one Solar radius of the centre of the white dwarf\footnote{The exceptions include WD 0806-661b 
\citep{luhetal2011} -- a planet orbiting a white dwarf at several thousand au -- and PSR B1620-26b 
\citep{sigetal2003} -- a circumbinary planet orbiting both a white dwarf and pulsar.}. This knowledge 
comes in three flavours, as described below, from: (i) atmospheric metal pollution, (ii) planetary debris discs, 
and (iii) active exo-minor planets.

\subsection{White dwarf planetary systems}

Metal pollution in white dwarf atmospheres refers to the detection of spectral lines which
cannot be intrinsic to the star \citep{schatzman1958,altetal2010,koester2013} and cannot 
arise from the interstellar medium \citep{aanetal1993,frietal2004,jura2006,kilred2007,faretal2010}.
These lines instead indicate metal-rich planetary debris which has been deposited \citep{wyaetal2014,broetal2017} 
in the atmosphere. Over a thousand metal-polluted white dwarfs are now known 
\citep{dufetal2007,kleetal2013,genetal2015,kepetal2015,kepetal2016,holetal2017}, and between
one-quarter and one-half of all observed white dwarfs are metal-polluted 
\citep{zucetal2003,zucetal2010,koeetal2014}. Twenty different atmospheric metals have been detected,
and overall they resemble the bulk Earth in composition 
\citep{kleetal2010,gaeetal2012,juryou2014,wiletal2015,wiletal2016,xuetal2017,haretal2018,holetal2018}.
Nearly all metal pollution has been sought around single white dwarfs, although confident
detections can be made in wide binary systems containing a white dwarf \citep{veretal2018}.

Over 40 metal-polluted white dwarfs are also known to harbour circumstellar planetary debris 
discs \citep{farihi2016}. These discs are compact (could fit within the Sun) and predominantly, 
unlike protoplanetary discs, are disorderly. White dwarf debris discs are variable \citep{xujur2014}, 
eccentric and wispy \citep{manetal2016}, and have demonstrated evidence of morphological changes
on both decadal \citep[from][]{gaeetal2006} and yearly \citep{denetal2018} timescales. Their formation
most likely arises from the break-up of minor bodies which have veered into the white dwarf
Roche radius \citep{graetal1990,jura2003,debetal2012,beasok2013,veretal2014a,veretal2015a} in 
combination with the migration of rocky fragments \citep{donetal2010,veretal2015b} produced from 
radiative destruction and/or deformation on the giant branch phase \citep{veretal2014b,katz2018}. 
The subsequent evolution
of the discs is complex and involves an interplay of gas and 
dust \citep{rafikov2011a,rafikov2011b,metetal2012,kenbro2017a,kenbro2017b,mirraf2018}.

The minor planet break-up scenario has been evidenced directly, and can in fact be seen on
a daily basis around the white dwarf WD 1145+017 when in season. \cite{vanetal2015} discovered
photometric transit curves suggesting that one or more exo-asteroids are in the process of
shedding dust and/or gas around this star. This seminal discovery has lead to over 20 more
dedicated papers about this system.  The object orbiting this white dwarf is the 
only known exo-minor planet, and its dynamical origin remains an open question.

\subsection{Linking with full-lifetime simulations}

Despite this uncertainty, major planets are thought to perturb the minor planets
into the white dwarf Roche radius\footnote{Second generation formation of minor
planets around white dwarfs remains a possibility \citep{schdre2014,voletal2014,hogetal2018,vanetal2018}.}.
Therefore, understanding how planets evolve over time and stellar phase is crucial, as
is determining when gravitational instabilities occur. Although the majority of literature
on planet-planet scattering focuses on the main sequence, we are instead interested in
the later phases of stellar evolution.

Efforts to study these scenarios have increased markedly over the last decade \citep{veras2016a}. 
Post-main-sequence investigations feature models of single-star exoplanetary systems with an asteroid belt
\citep{donetal2010,bonetal2011,debetal2012,frehan2014,musetal2018,smaetal2018}, 
single-star exoplanetary systems without
an asteroid belt \citep{debsig2002,veretal2013a,voyetal2013,musetal2014,vergae2015,veretal2016,payetal2016,payetal2017}, and multiple-star exoplanetary systems 
\citep{kraper2012,vertou2012,musetal2013,portegieszwart2013,bonver2015,hampor2016,kosetal2016,petmun2017,steetal2017,veretal2017a,veretal2017b,steetal2018}.

One common theme amongst nearly all of the listed papers above is that they either make an assumption of coplanarity, near-coplanarity (mutual inclinations smaller than a couple of degrees), or focus on the high inclination case, where the Kozai-Lidov mechanism acts 
\citep{kozai1962,lidov1962,naoz2016}.
Therefore, there is a need to understand how good or bad the assumption of near-coplanarity is with regard to
stability across multiple phases of stellar evolution, and explore the full mutual inclination range from $0^{\circ}-40^{\circ}$. As mentioned earlier, our knowledge of the mutual inclination
of the multiple exoplanets in a given exosystem is lacking, and being prepared for revelations from {\it Gaia} will be
beneficial.

\subsection{This paper}

Therefore, this straightforward paper simply aims to assess, qualitatively, the reliability of the near-coplanar assumption, 
and hence whether it can continue to be applied with confidence to future modelling efforts and exploratory simulations. We do not
attempt to cover a wide range of phase space nor provide detailed dynamical analyses. Instead, we have performed a large
ensemble ($\sim 10^4$) of CPU-intensive full-lifetime simulations for the simplest case of two circular, inclined planets orbiting a 
single star that loses mass isotropically (itself a good assumption; \citealt*{veretal2013b}). This paper then acts as
a basic extension to both \cite{verarm2004}, which considered the dynamics of two planets on circular inclined orbits along the 
main sequence, and \cite{veretal2013a}, which investigated two circular coplanar planets through all phases of stellar evolution. 
We review our knowledge
of stability boundaries in Section 2; Section 3 details
our simulations, Section 4 characterizes our results, and Section 5 summarizes our conclusions.

\section{Stability boundaries}

As in any three-body system, there exist analytical criteria to determine the critical separations beyond 
which the mutual orbits will never cross \citep{georgakarakos2008}. These criteria describe {\it Hill stable} 
configurations, and rely on the allowed parameter space regions from energy and angular momentum considerations
\citep[e.g.][]{marboz1982}. For systems of one star and two equal-mass planets on circular orbits, this separation $\Delta$ can be expressed explicitly as a function of mutual inclination $i_{\rm mut}$ \citep{verarm2004} as

\[
\Delta \approx \epsilon + \eta \sqrt{ 
\left(4 + \frac{\cos^2{i_{\rm mut}}}{2}  \right)
\left(\epsilon + \chi \eta \mu^{2/3}  \right)
} 
\]

\begin{equation}
\ \ \ \ 
+ \chi \eta \mu^{2/3} - 3 \eta^2 \mu \sqrt{
\frac{4 + \frac{\cos^2{i_{\rm mut}}}{2}}
{\epsilon + \chi \eta \mu^{2/3}}
}
\label{Hillinc}
\end{equation}

\noindent{}where

\begin{equation}
\epsilon = 2 + \cos^2{i_{\rm mut}} - \cos{i_{\rm mut}} \sqrt{8 + \cos^2{i_{\rm mut}} }
,
\end{equation}

\begin{equation}
\eta = 1 - \frac{\cos{i_{\rm mut}}}{\sqrt{8 + \cos^2{i_{\rm mut}}}}
,
\end{equation}

\begin{equation}
\chi = 3 \cdot 2^{1/3} \cdot 3^{1/3}
,
\end{equation}

\noindent{}and $\mu$ is the planet-star mass ratio. For arbitrarily inclined and eccentric orbits,
the criteria may be written as a set of equations to be solved both explicitly and implicitly 
\citep{donnison2006,donnison2009,donnison2011} and generalized to include secular and evection processes 
\citep{grietal2017} and the angular momentum deficit \citep{petetal2018}. Although useful, 
Hill stability criteria are not exact
\citep[see Section 6.5 of][]{veretal2013a} and become progressively conservative and less useful as the mutual
inclination increases, particularly near mean motion commensurabilities \citep{verarm2004}. Hence, one 
may instead appeal to stability boundaries approximated by numerical simulations 
\citep[e.g.][]{georgakarakos2013,petrovich2015}.

Because Hill stability refers only to the situation where orbits cannot cross, it does not take into
account situations where the inner planet collides with the star or the outer planet escapes the system.
Hence, Hill stability does not represent a proxy for global stability, known as Lagrange instability
\citep{bargre2006,bargre2007,rayetal2009,vermus2013,marzari2014}. Therefore, Hill stable systems may
be globally unstable. Similarly, Hill unstable systems may be globally stable. In this latter case, a 
mean motion resonance could provide forcing which protects the system.

Regardless of the stability prescription one uses, as the parent star leaves the main sequence, it will 
shed mass and expand the orbits of accompanying planets
\citep{omarov1962,hadjidemetriou1963,veretal2011}. The stability boundary varies at a different pace 
than the semimajor axis ratio of the planets, 
potentially triggering instability \citep{debsig2002}. What remains unclear is how the stability boundaries
change in the non-coplanar case, which is the focus of this paper.

In the near-coplanar case, the instability may occur during the giant branch phase, but has been shown to 
predominantly be delayed
until the star has become a white dwarf \citep{veretal2013a}. The implications for white dwarf
disc creation and pollution are crucial, as described in Section 1. When a planetary orbital period
exceeds the stellar mass loss timescale, then the planet's orbital eccentricity can change appreciably  
\citep{veretal2011,adaetal2013}, providing for 
an additional factor in stability boundary changes \citep{voyetal2013}. However, eccentricity variations
due to mass loss need not be considered here, as the majority of known white dwarf planetary system hosts
arise from A and F stars \citep{treetal2016}. These stars have peak mass loss timescales which are several orders
of magnitude greater than planetary orbital periods, for the vast majority of known planets.

\section{Simulation setup}

\subsection{Initial condition choices}

Indeed, observationally motivated by these A-type and F-type progenitors, we adopted
main sequence stellar masses of $1.5M_{\odot}$ and $2.0M_{\odot}$ for 
the majority of our simulations, which ran for 14 Gyr. 

We also included simulations with higher stellar
masses $2.5M_{\odot}$, $3.0M_{\odot}$ and $5.0M_{\odot}$ in order to showcase potentially 
extreme cases of dynamical evolution. These latter two classes of simulations also 
run more quickly than
those for A and F stars because of their shorter main sequence lifetimes; for details
on these differences, see \cite{veretal2013a}.
We note that in \cite{veretal2013a}, computational limitations prevented us from simulating
host star masses lower than $3.0M_{\odot}$; in the intervening five years, these
capabilities have improved\footnote{Simulating a stellar host with a mass as low as 1.0 $M_{\odot}$
over $\approx 10^{10}$ yr remains a CPU-intensive challenge, particularly when using the newly-released
and more accurate (but slower) code revealed in \cite{musetal2018}. One partial way to circumvent
this issue is to begin simulations towards the end of the main sequence \citep{veras2016b}.}.
In \cite{verarm2004}, main sequence simulations of inclined planets were run for just 
2 Myr, a full four orders of magnitude shorter than the simulations considered here!

As opposed to both \cite{verarm2004}
and \cite{veretal2013a}, here we sampled initial conditions for the planets in a Monte Carlo fashion
within particular parameter ranges. Doing so allowed us to add to our sample size
at will, which was useful given the computational demands of the simulations.
All our simulations featured two planets with initial mutual inclinations between $1^{\circ}$ and
$40^{\circ}$, because Kozai-Lidov oscillations may occur for inclinations higher than
$\arccos{(\sqrt{3/5})} \approx 39^{\circ}$. The initial inner planet semimajor axis was set
at 10 au, which is a realistic value and a high-enough value to enable us to complete
$10^4$ simulations. The initial outer planet semimajor axis was chosen randomly from
a uniform distribution within a range that would showcase both stability and instability.
The planets' true anomalies were chosen randomly from a uniform distribution 
across their entire possible ranges. In a small fraction of cases where we explored non-circular orbits,
a planet's initial longitude of ascending node was fixed at $0^{\circ}$ but its argument of pericentre
was chosen randomly from a uniform distribution across all possible values.

\begin{figure*}
\centerline{\textcolor{blue}{\bf \Large TWO 1.0 JUPITER-MASS PLANETS}}
\vspace{0.5cm}
\centerline{
\includegraphics[width=9.3cm]{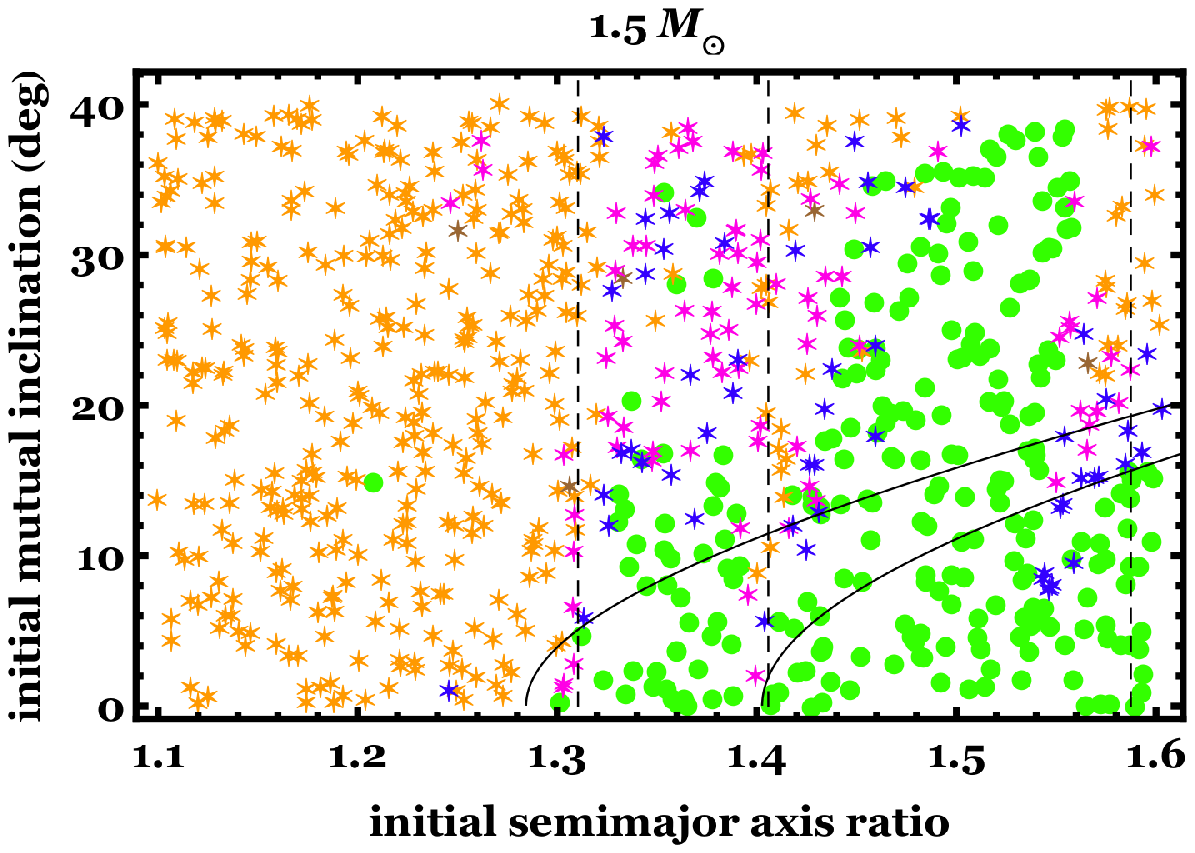}
\ \ \ \ \ \ \ \
\includegraphics[width=9.3cm]{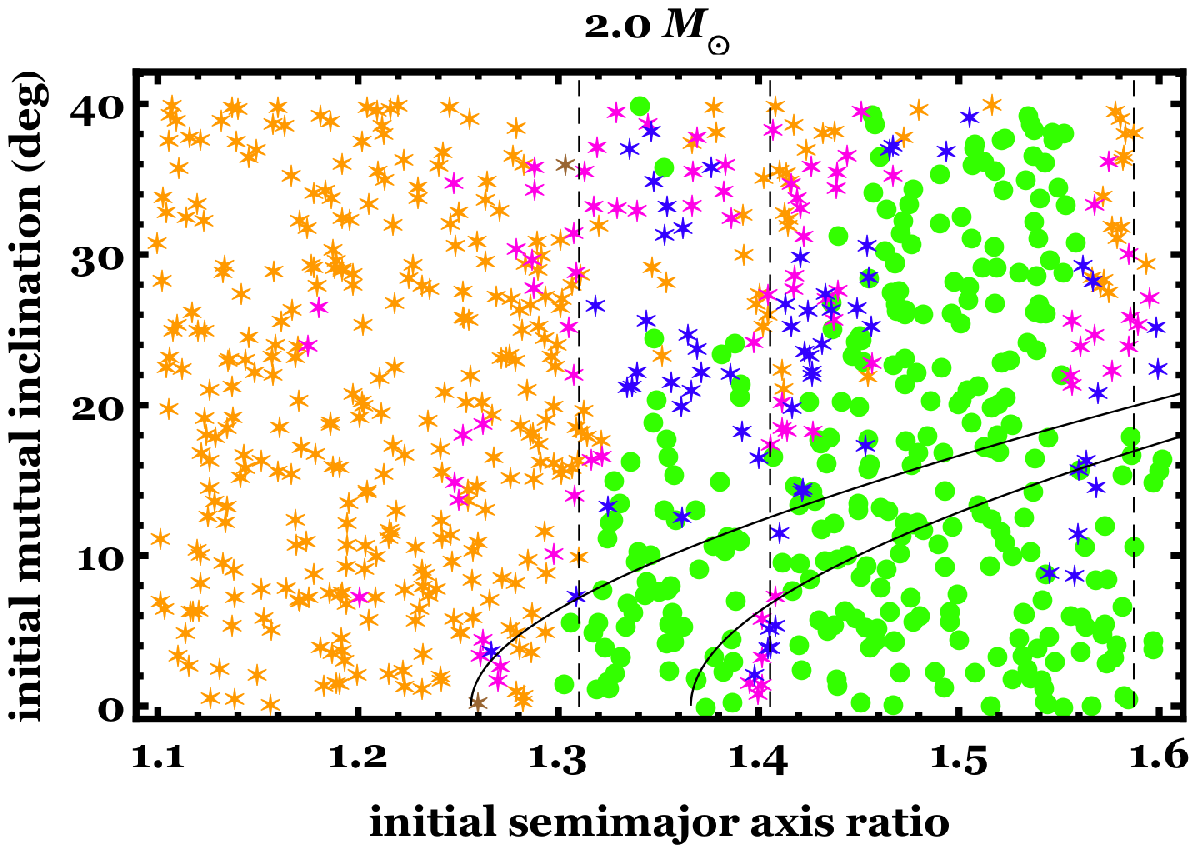}
}
\vspace{0.5cm}
\centerline{
\includegraphics[width=9.3cm]{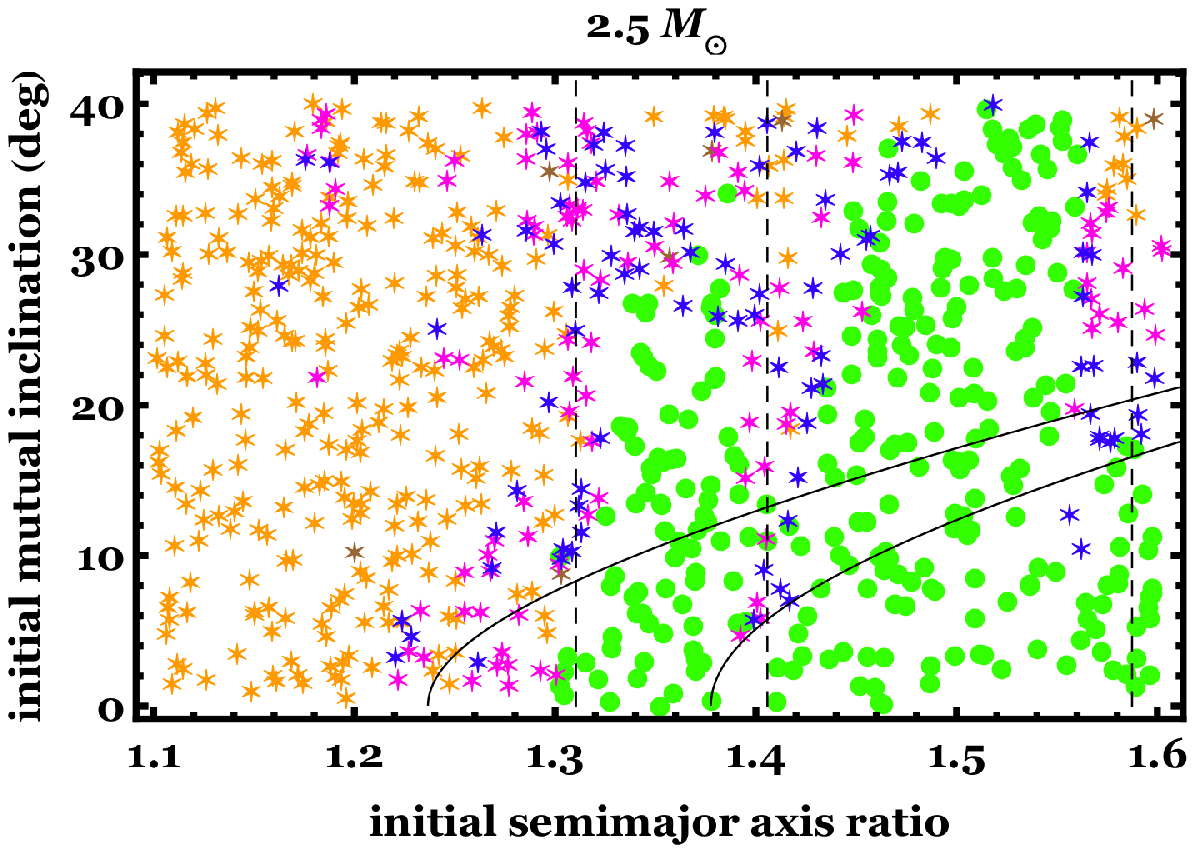}
\ \ \ \ \ \ \ \
\includegraphics[width=9.3cm]{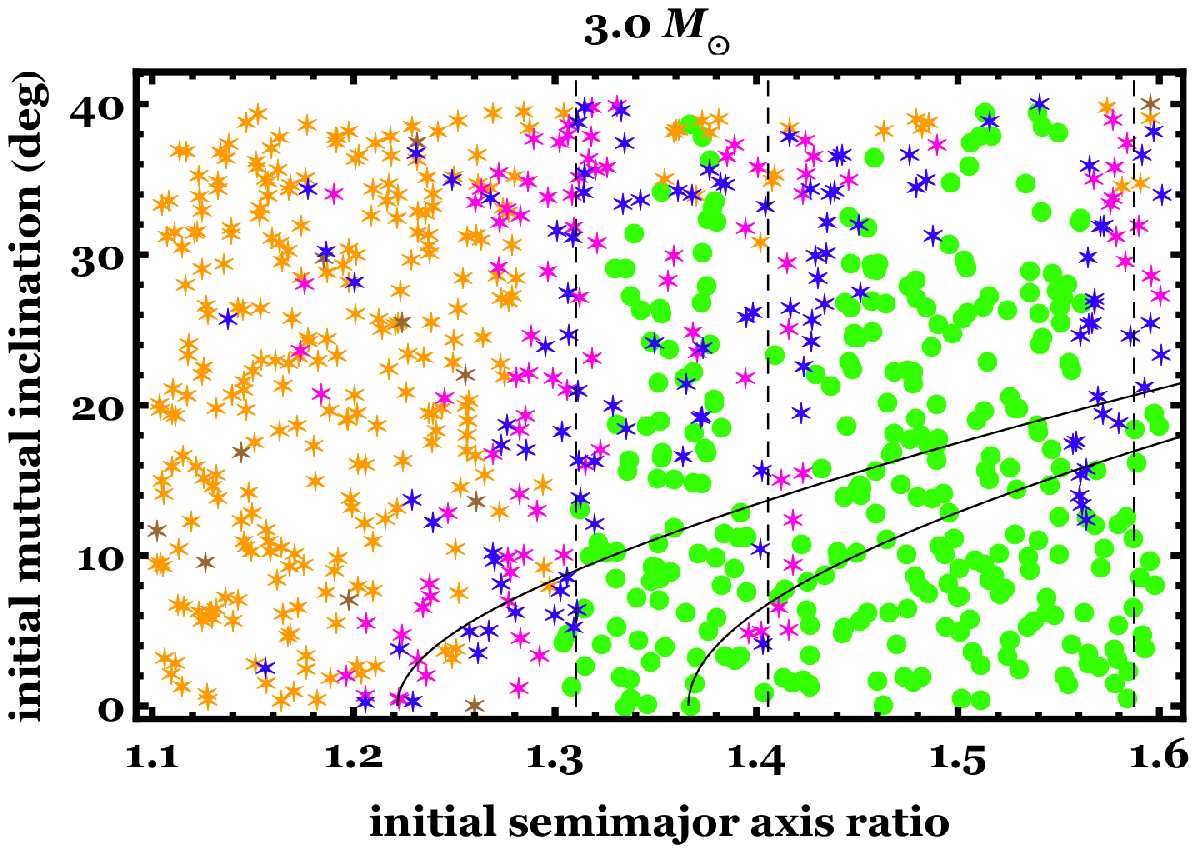}
}
\vspace{0.5cm}
\centerline{
\includegraphics[width=9.3cm]{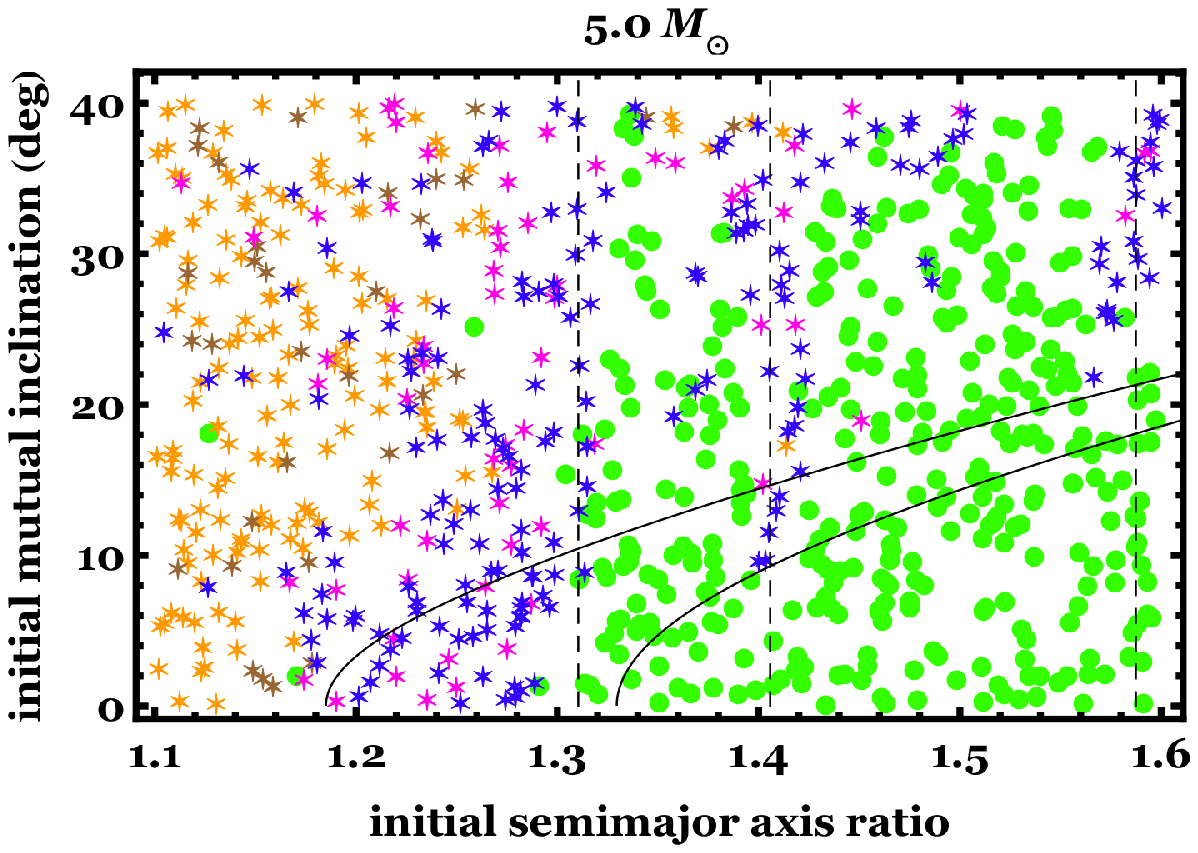}
\ \ \ \ \ \ \ \
\includegraphics[width=8.0cm]{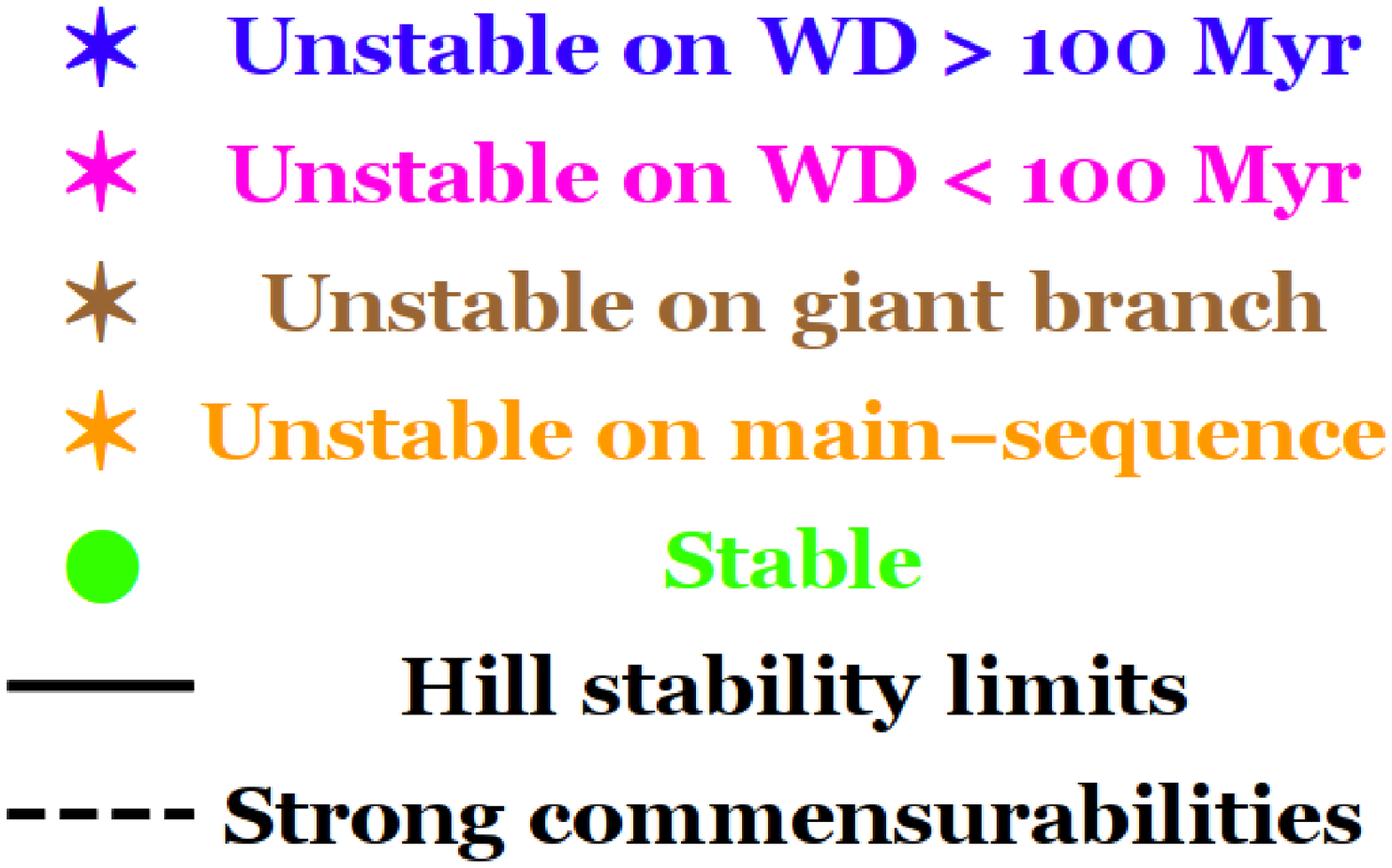}
}
\caption{
Full-lifetime (14 Gyr) simulations of two $1.0 M_{\rm Jup}$ planets on initially
circular orbits around a star with the progenitor main sequence mass stated
in the plot title.
Dots (which are green) indicate stable simulations, whereas star symbols
indicate unstable simulations. Orange and brown stars respectively represent
instability on the main sequence and giant branch phases of stellar evolution.
Purple and blue stars represent instability on the white dwarf phase
when the white dwarf is respectively younger or older than 100 Myr. Overplotted as solid
black lines is the Hill
stability limit (equation \ref{Hillinc}) for the main sequence (upper curve) and
white dwarf phase (lower curve). Overplotted as vertical dashed lines are,
from left to right, the mean motion commensurabilities $3$:$2$, $5$:$2$ 
and $2$:$1$. The plots illustrate that despite the stability dependencies on
initial parameters, (i) late instability along the white dwarf phase occurs
throughout the sampled inclination range, and (ii) the assumption of near-coplanarity
would be most adequate as a representation for ensemble global stability studies 
when the planets are away from strong mean-motion commensurabilities or within
the Hill stability semimajor axes ratio limits (but external to global chaos from
resonance overlap boundary, here near the $3$:$2$ commensurability).
}
\label{JJcirc}
\end{figure*}

\begin{figure*}
\centerline{\textcolor{blue}{\bf \Large TWO 1.0 EARTH-MASS PLANETS}}
\vspace{0.5cm}
\centerline{
\includegraphics[width=9.3cm]{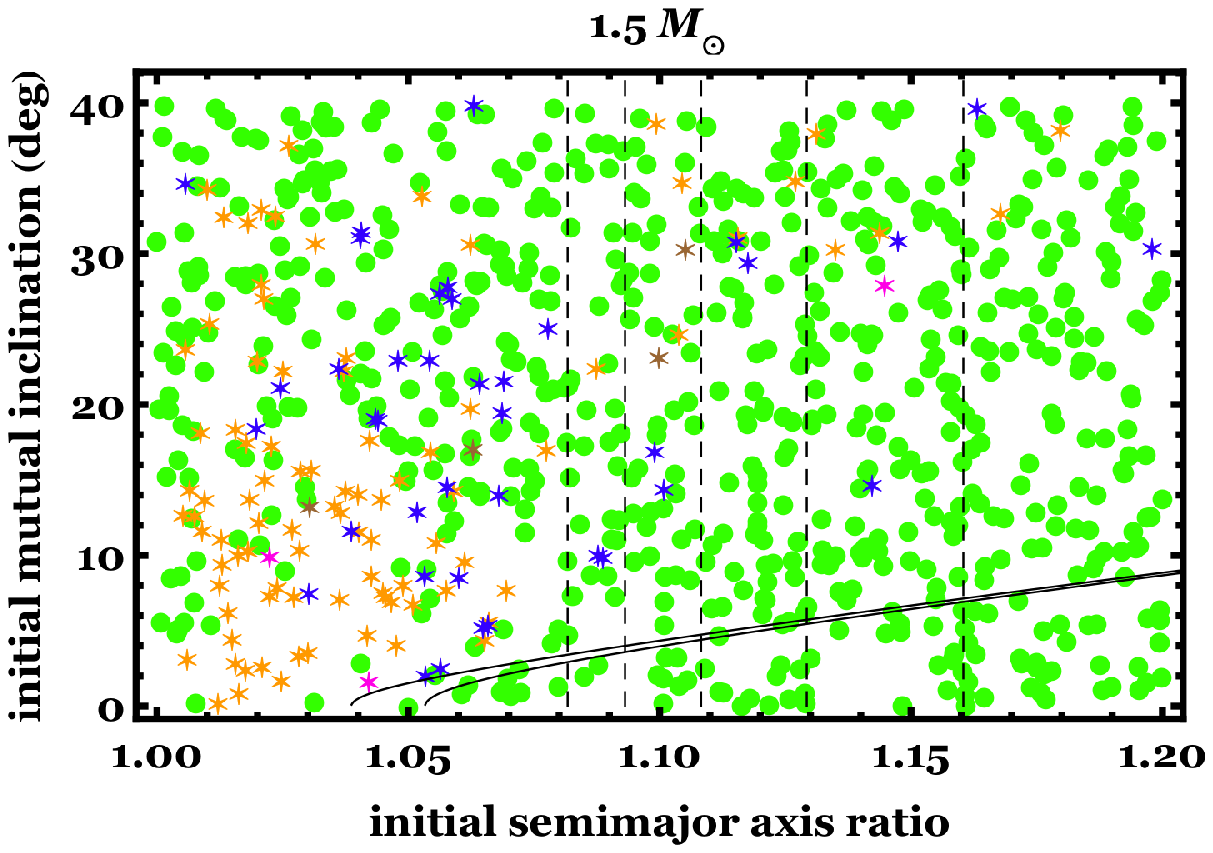}
\ \ \ \ \ \ \ \
\includegraphics[width=9.3cm]{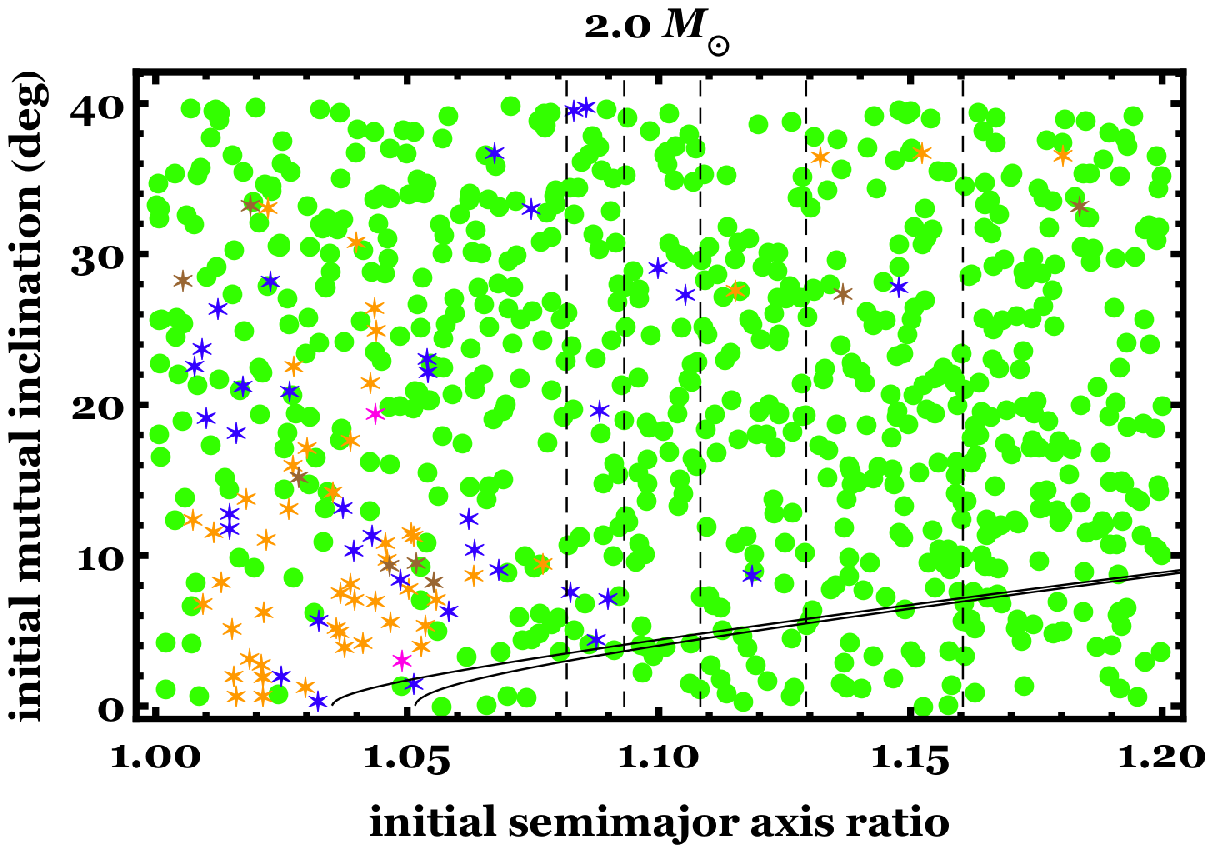}
}
\vspace{0.5cm}
\centerline{
\includegraphics[width=9.3cm]{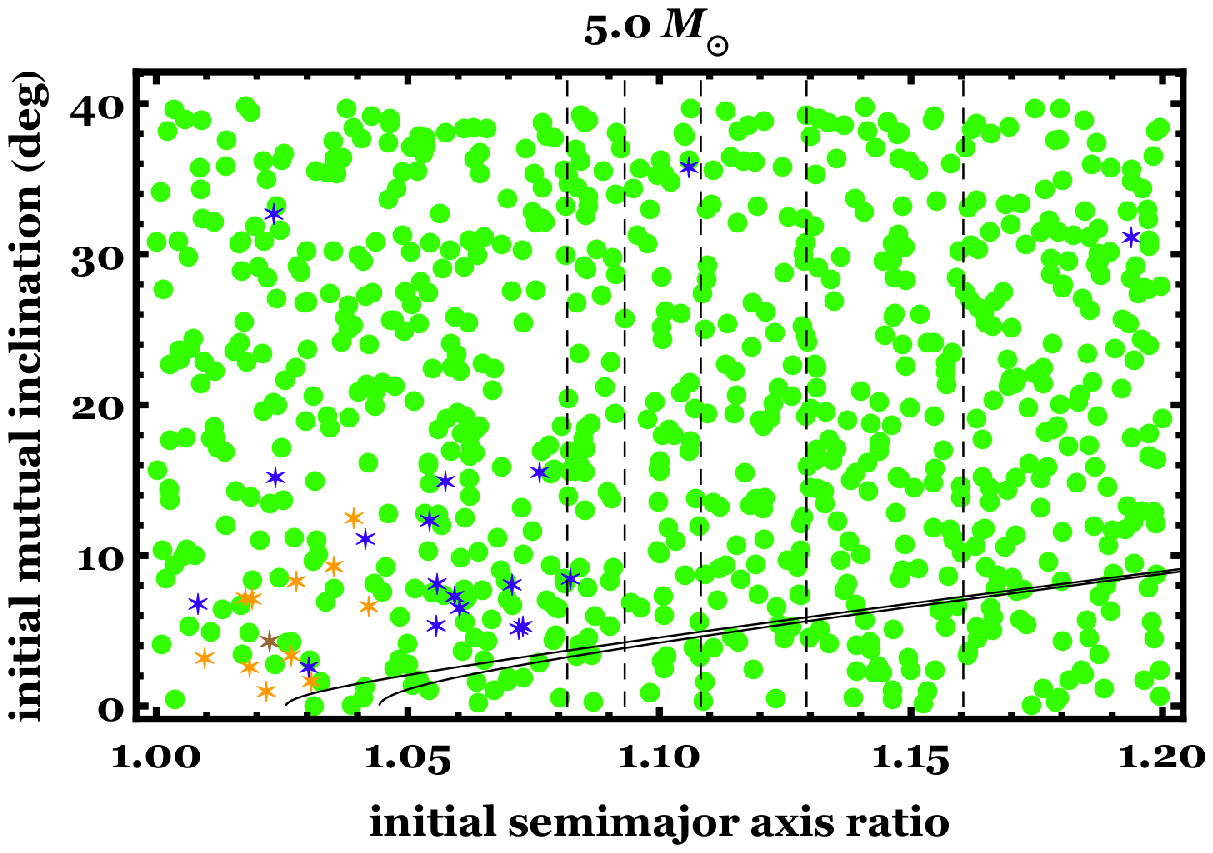}
}
\caption{
Same as Fig. \ref{JJcirc}, except for Earth-mass planets and highlighted mean motion 
commensurabilities of, from left to right, $9$:$8$, $8$:$7$, $7$:$6$, $6$:$5$ and $5$:$4$.
In this case, the near-coplanar assumption would excel, except within the Hill stability
semimajor axis ratio limit. Late instability on the white dwarf phase can occur for any   
initial mutual inclination, although that represents a rare outcome for these phase portraits. 
}
\label{EEcirc}
\end{figure*}

\begin{figure*}
\centerline{\textcolor{blue}{\bf \Large A 1.0 JUPITER-MASS PLANET (INNER) AND}}
\centerline{\textcolor{blue}{\bf \Large A 1.0 EARTH-MASS PLANET (OUTER)}}
\vspace{0.5cm}
\centerline{
\includegraphics[width=9.3cm]{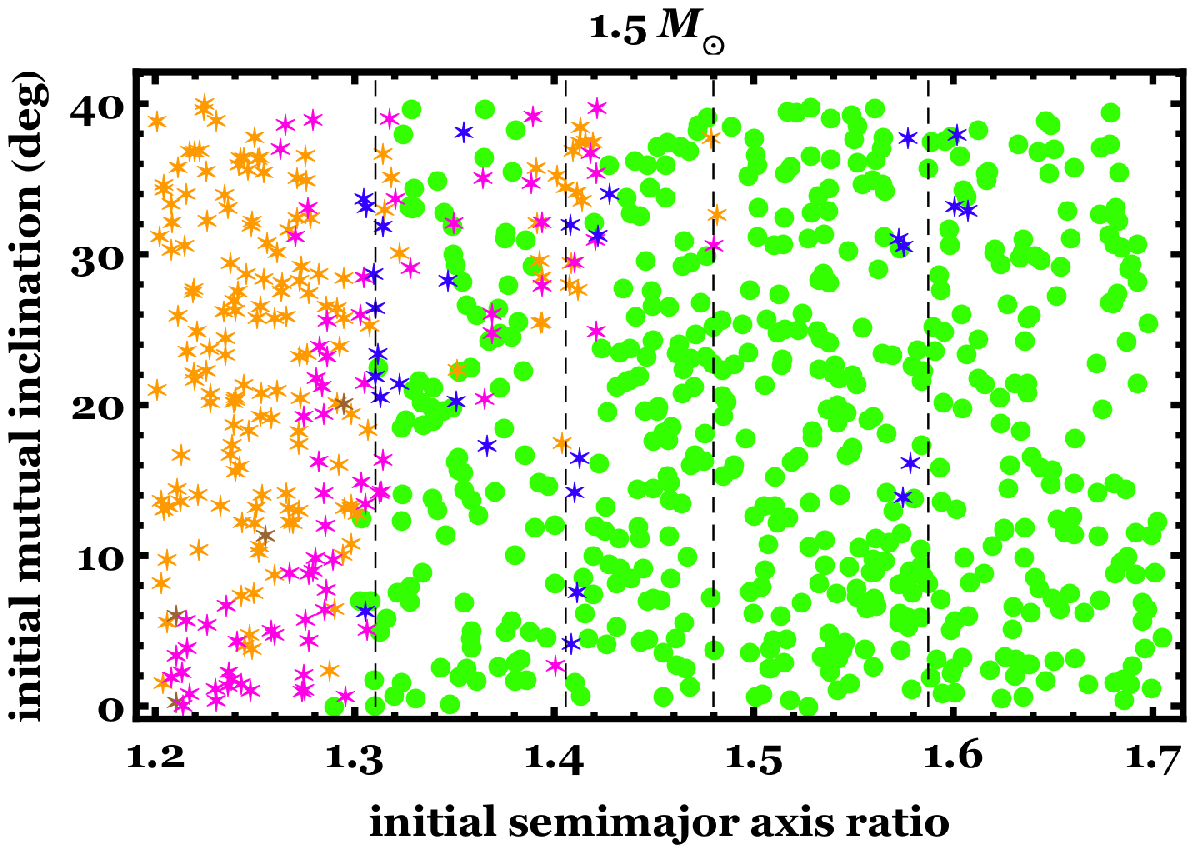}
\ \ \ \ \ \ \ \
\includegraphics[width=9.3cm]{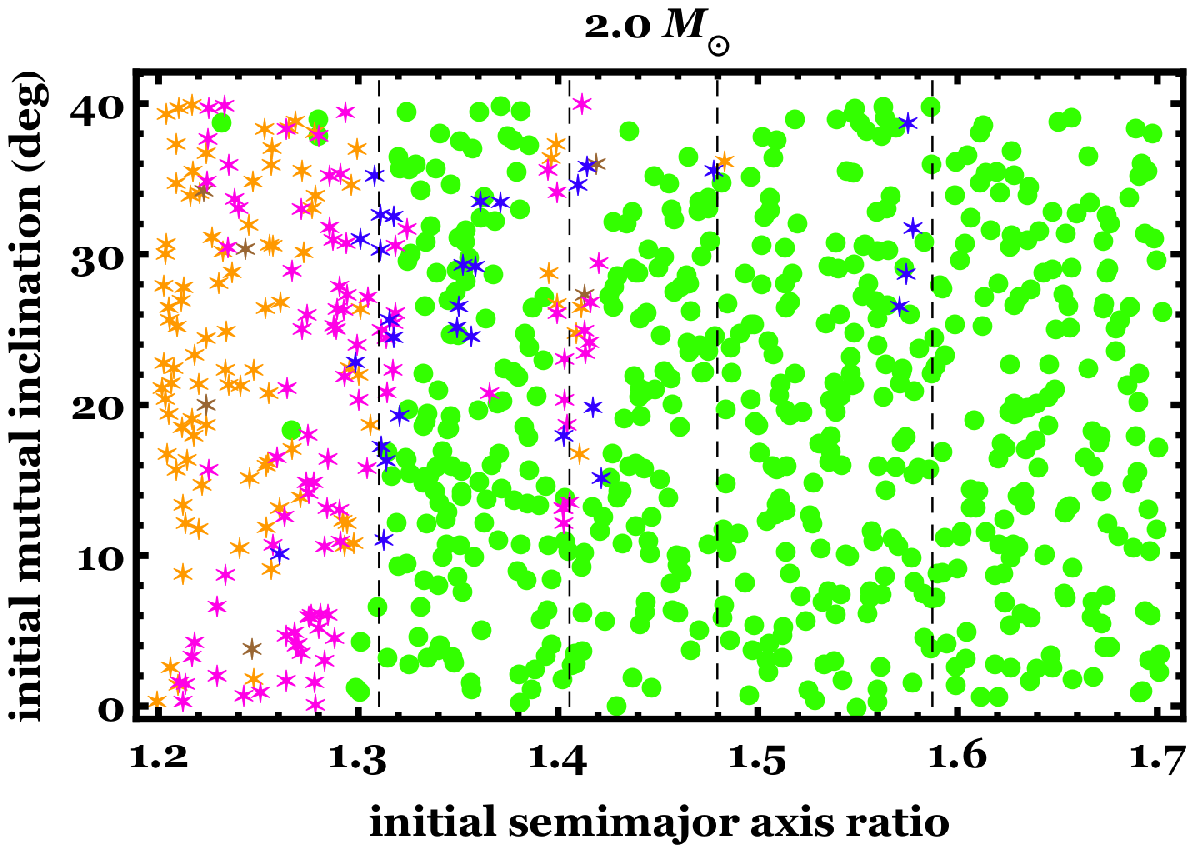}
}
\vspace{0.5cm}
\centerline{
\includegraphics[width=9.3cm]{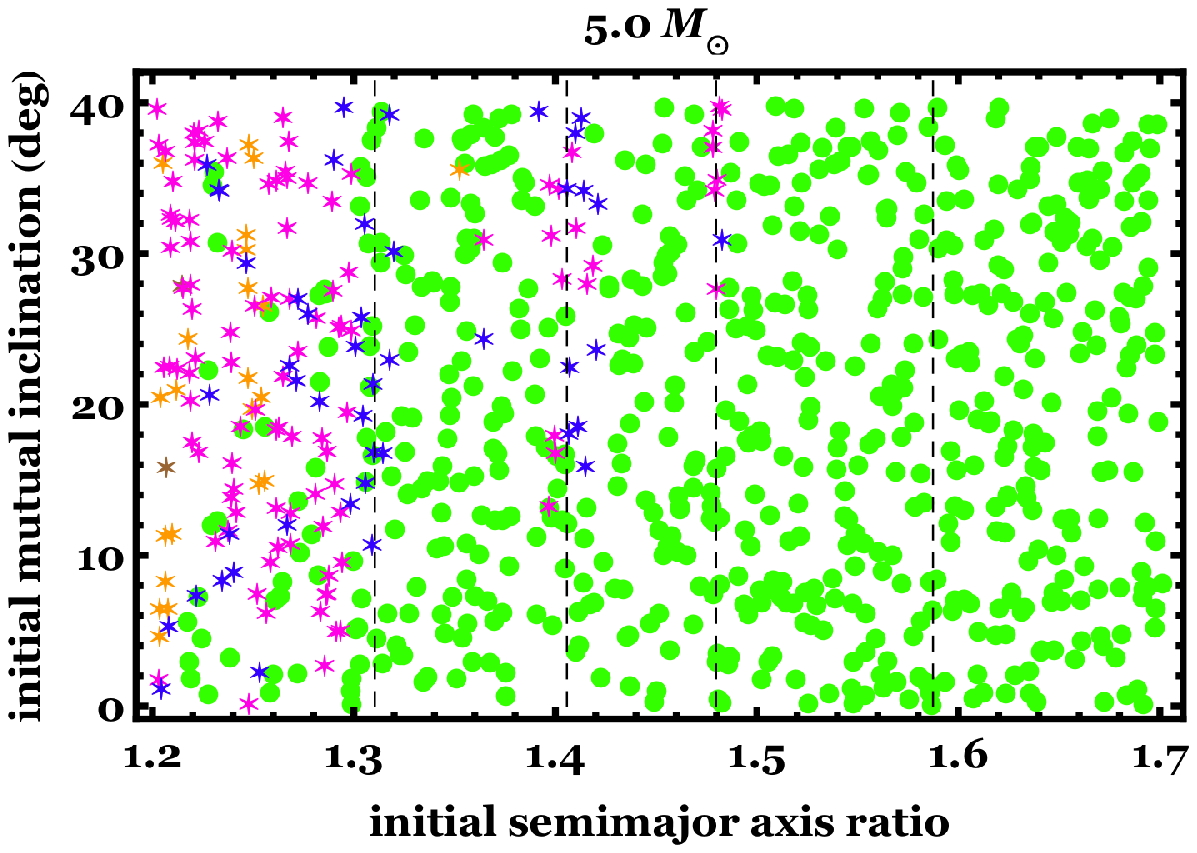}
}
\caption{
Same as Fig. \ref{EEcirc}, except for two unequal-mass planets (an inner Jupiter and an outer Earth) 
and highlighted mean motion commensurabilities
of, from left to right, $3$:$2$, $5$:$3$, $9$:$5$ and $2$:$1$. Here, the near-coplanar assumption 
would remain robust anywhere in semimajor axis ratio phase space except around mean motion commensurabilities.
Late instability on the white dwarf phase is more strongly tied to the highlighted commensurabilities
than in the other plots, but can still occur for any initial mutual inclination.
}
\label{JEcirc}
\end{figure*}

In order to facilitate comparison with the outputs from \cite{verarm2004} and \cite{veretal2013a},
we chose two Jovian-mass planets on circular orbits for our fiducial set of simulations.  
We achieved the highest resolution for these simulations, completing 800 simulations for each
of the five sampled stellar masses (Fig. \ref{JJcirc}). We also explored other two-planet 
scenarios: (i) given a pair of Earth-mass planets, we ran 800 simulations for each of 
three different stellar masses (Fig. \ref{EEcirc}), and (ii) given a planet pair including 
one inner Jovian-mass planet and one outer Earth-mass planet, we ran 800 simulations for each of three
different stellar masses (Fig. \ref{JEcirc}). Finally, we ran a low-resolution ensemble
of 400 simulations for a main-sequence stellar mass of $5.0M_{\odot}$ and a pair of two 
Jovian-mass planets, where the inner one was on an initially circular orbit, and the outer 
one's orbit had an initial eccentricity of 0.3 (Fig. \ref{JJecc}).

\subsection{Simulator}

We used the newly-released planetary evolution code used in \cite{musetal2018} and described
in detail in A. Mustill et al. (in prep). This code executes the RADAU integrator to propagate planets
forward in time, and is a modified version of the code used
in \cite{veretal2013a}, where the Bulirsch-Stoer integrator was adopted. That code is itself a
modified version of the {\it Mercury} integration package \citep{chambers1999}. Stellar evolution is
incorporated by interpolating output from the {\tt SSE} code \citep{huretal2000} into the timesteps of the
planetary propagator. 

The {\tt SSE} code was run with its default parameters. These include a stellar metallicity of $Z=0.02$,
a value of 0.5 for the Reimers mass loss coefficient along the giant branch phase, and the inclusion of a superwind along the asymptotic giant branch phase, as prescribed by \cite{vaswoo1993}. Although the choice of Reimers mass loss coefficient may qualitatively change evolutionary pathways for Solar-mass stars 
\citep{schcun2005,verwya2012,veras2016b}, its importance wanes for higher-mass stars like the ones considered here. In fact, as shown in Fig. 3 of \cite{veras2016a}, of the total mass eventually 
lost by a main-sequence star of mass $1.5 M_{\odot}$, less than 10 per cent of that loss will occur  
along the red giant phase; for a star with a main-sequence mass of $2.0 M_{\odot}$, the fraction
lost along the giant phase is on the order of just $0.1$ per cent.

In the planetary evolution code, the improvement afforded by the RADAU 
integrator over the Bulirsch-Stoer integrator can be significant
(see Appendix A of \citealt*{musetal2018}), but comes at a price in computational speed. 
In order to achieve a balance of accuracy and speed, for all our simulations, we adopted a 
tolerance of $10^{-11}$. According to Fig. A1 of \cite{musetal2018}, this tolerance would allow
us to accurately track mean anomaly evolution to within about a dozen degrees over $\approx$ 5 Gyr.
Such accuracy -- which is much higher than what was achieved in \cite{veretal2013a} -- aids in the 
ensemble study of stability performed here, but is perhaps still too low to make detailed conclusions
about resonance retention, capture and expulsion across different phases of stellar evolution.

\section{Results and interpretation}

\subsection{Summary}

We summarise our results with plots of initial mutual inclination versus initial semimajor
axis ratio, and use different symbols and colours to indicate stability or instability
at different phases (Figs. \ref{JJcirc}-\ref{JJecc}). In all plots, we ran 800 simulations,
except for Fig. \ref{JJecc}, where we ran 400 simulations. We were more interested in
vertical trends on the plots rather
than horizontal trends, because the latter have already been investigated in
detail \citep{veretal2013a}.
These plots demonstrate that the vertical trends are weak or nonexistent. Hence, 
we conclude that instability is largely independent of non-Kozai mutual inclination
except close to mean motion commensurabilities.

\subsection{Details}

Now we provide some details of the plots, which illustrate instability at various times, and distinguish
stable systems from unstable systems. By the term ``instability'' we are referring to a disruption
in the system: instantiated either by planetary escape from the system, or with a collision 
(between both planets or with the star and planet)\footnote{We define the escape boundary as the Hill ellipsoid
surrounding the star, as in previous studies \citep{shaetal2014,veretal2014c,veras2016b}. 
``Hill ellipsoid'' here refers to the region within the Galactic Disc which is 8 kpc away from 
the Galactic centre where objects are gravitationally attracted to the star \citep{vereva2013,veretal2014d}.}.  

The meaning of the colours and symbols are as follows.
Green dots indicate stable systems. Unstable
systems, denoted by star symbols, are distinguished by colour: the abundant orange symbols indicate main 
sequence instability, the rare brown symbols indicate instability on the giant branch phase,
and the purple and blue symbols represent instability on the white dwarf phase, for, respectively,
white dwarf ages (known as ``cooling ages'') less and greater than 100 Myr. 

This last division
was motivated by giant branch mass loss having a well-documented effect of predominantly delaying
instability until just after the white dwarf is 
born; later instability does occurs, but decreases in frequency with time
\citep{veretal2013a,musetal2014,vergae2015,veretal2016,musetal2018}. 
The later instabilities are particularly important as a way to explain the white dwarf pollution
observed in white dwarfs which are Gyrs old \citep{holetal2018}. The scatter in blue symbols
on the plots demonstrate that late instability along the white dwarf phase may occur at
any value of the initial mutual inclination, and without a strong preference for low or high values.

This scatter emphasises the sensitivity of the stability of planetary systems to initial conditions.
The sensitivity to the initial semimajor axis ratio is seen most starkly at the value where the bulk
of the main-sequence instability (orange star symbols) ends. This value lies roughly at the location of the Hill
stability boundary for circular orbits ($i_{\rm mut} = 0^{\circ}$ in equation \ref{Hillinc}). We have
plotted equation (\ref{Hillinc}) in Figs. \ref{JJcirc} and \ref{EEcirc} as black curves, both for
the main sequence mass (upper curve) and white dwarf mass (lower curve). The first figure illustrates 
that the inclination-dependence of Hill stability for both main sequence and white dwarf systems 
with Jupiter-mass planets roughly mirrors the trend that instability increases with higher initial
mutual inclination.

However, the greater dependence on stability arises from proximity to a strong mean motion resonance.
Hence, we have overlayed dashed 
vertical lines on the plots indicating the locations of commensurabilities of interest due to their strength
and location. We chose our commensurabilities in order to (i) demonstrate clear connections between commensurability 
and instability, and (ii) in cases of near-ubiquitous stability throughout the phase space (Fig. \ref{EEcirc}), 
to demonstrate that even the strongest commensurabilities may not affect system stability. As the accuracy
and output frequency of our long-term simulations are not quite at a level where a proper resonant 
analysis could be performed,
we are content here to just show proximity to commensurability. Overall, the commensurability locations
correspond to strips of instability (star symbols) amidst a sea of stability (green dots).

This trend is clear for the Jupiters in Fig. \ref{JJcirc}, where from left to right the $3$:$2$, $5$:$2$ 
and $2$:$1$ commensurabilities are plotted. These commensurabilities were computed strictly from orbital
period ratios only, without incorporating the time evolution of the longitudes of ascending node, nor
the planet masses. Note that regardless of the fact that inclination-based resonances must be of at least
order two, both the $3$:$2$ and $2$:$1$ commensurabilities clearly correlate with instability. One
potential reason is that around these locations, the resonant angles associated with the $6$:$4$,
$9$:$6$, $4$:$2$ or $6$:$3$ commensurabilities could be librating \citep{veras2007}. Further,
the $3$:$2$ commensurability lies close to
the circular Hill stable limit, and both are related through resonant overlap 
\citep{wisdom1980,mardling2008,quillen2011,decetal2013,rametal2015,hadlit2018}.

In fact, for the Earths in Fig. \ref{EEcirc}, the lack of correlation between the green dots (stable systems) and the first-order mean motion commensurabilities (plotted from left to right are the $9$:$8$, $8$:$7$, $7$:$6$, $6$:$5$ and $5$:$4$ commensurabilities)
illustrates either that (i) resonant overlap has washed away any observable trend
between nominal commensurability locations and
instabilities, or (ii) simply that the associated first-order resonant librating angles
cannot be inclination-based. In contrast, the trend is strongest in Fig. \ref{JEcirc}
(from left to right,
the $3$:$2$, $5$:$3$, $9$:$5$ and $2$:$1$ commensurabilities are drawn), which features one Jupiter
and one Earth. Unequal mass objects allow for stronger signatures of resonant capture because 
in this case resonant angles are more likely to librate, as they could include some terms which may
be neglected. The most striking trend is the narrowly-confined band of instability for
the $5.0 M_{\odot}$ main sequence
progenitor. As a fourth-order resonance, and the weakest highlighted on the figure, the libration
width in initial semimajor axis ratio space would be the smallest.

\begin{figure}
\centerline{\textcolor{blue}{\bf \Large TWO 1.0 JUPITER-MASS PLANETS,}} 
\centerline{\textcolor{blue}{\bf \Large WITH THE OUTER ONE ECCENTRIC}}
\vspace{0.5cm}
\centerline{
\includegraphics[width=9.3cm]{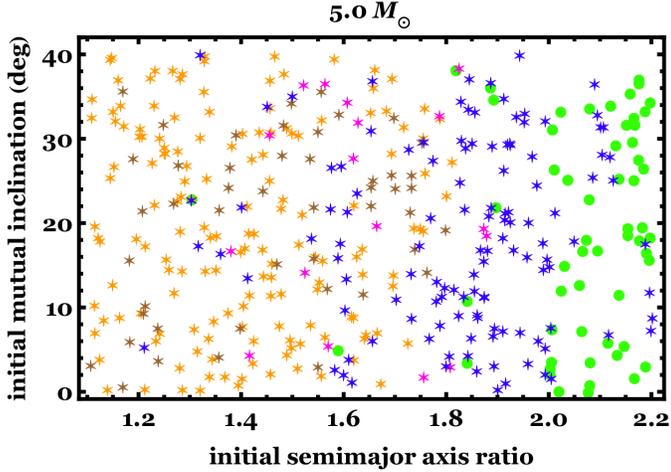}
}
\caption{
Same as Fig. \ref{JJcirc}, except the outer planet's orbit is given an initial eccentricity of 0.3.
Although the horizontal scatter in the stability boundaries is more evident in this plot, it
reinforces the main conclusions of the paper.
}
\label{JJecc}
\end{figure}

Despite these trends, the plots demonstrate that non-Kozai mutual inclination represents a weak distinguishing 
factor in the stability of full-lifetime planetary systems. Adding an additional degree of freedom such as
eccentricity complicates the dependency further. As an example, consider Fig. \ref{JJecc}, where we set
the initial eccentricity of the outer planet to 0.3. As a result, the stability limits and commensurability 
locations change, yielding a fuzzier semimajor axis ratio boundary between stability and instability. 
Nevertheless, no vertical trends are apparent, reinforcing the conclusions of this study.

Ensemble studies like this one could always benefit from a finer sampling of phase space 
and greater integrator accuracy. \cite{georgakarakos2013} created a detailed phase portrait
in mass and mutual inclination space of stability for two planets orbiting a main sequence
star. By integrating some systems for up to $10^7$ orbits of the outer planet, he 
showed that mutual 
inclination does affect
the stability boundary across all mutual inclinations from $0^{\circ}$ to $180^{\circ}$, but 
only weakly when $i_{\rm mut} \approx 0^{\circ}-40^{\circ}$. Here, in our more broad study, we place 
all unstable main sequence systems in the same category, despite the different main sequence 
lifetimes of the sampled progenitor star masses. Future studies could involve not just a more expansive
sampling of the phase space, but also the inclusion of more planets, and an inclination-based
extension to the planet packing hypothesis \citep{chaetal1996,rayetal2009,fanmar2013,krasha2014,puwu2015,hwaetal2017}, 
particularly on the post-main-sequence \citep{vergae2015,veretal2016}. A different type of
extension might involve exploring how mutual inclination affects the generation of high eccentricities
(required to generate white dwarf metal pollution) from particular mean motion resonances
\citep{picetal2017}, secular resonances \citep{smaetal2018} and periodic orbits \citep{antver2016}.

\section{Summary}

The {\it Gaia} mission will, likely for the first time, (i) discover planets orbiting white dwarfs
at distances of a few au, and (ii) allow the mutual inclination of multiple planets to be measured
for an ensemble of systems. However, no dedicated exploration of the effects of non-Kozai mutual 
inclinations in multi-planet systems across all phases of stellar evolution is available. 
This paper has attempted to fill this gap, at least with an initial statistical
study, partly to help assess the goodness of the near-coplanar assumption which has predominately been applied until now.
We performed about $10^4$ computationally-intensive simulations of two-planet systems which ran for 14 Gyr across all 
phases of stellar evolution, and utilised an accurate RADAU code with an adopted tolerance of $10^{-11}$ 
(Mustill et al. 2018; A. Mustill in prep). We found that instabilities occurring around old white dwarfs
can be triggered from any initial mutual inclination between $0^{\circ}$ and $40^{\circ}$, and that
instability along
any post-main-sequence phase is a weak function of initial mutual inclination, but a strong function
of proximity to mean motion commensurability. Consequently, a measured mutual inclination alone
would not a reliable indicator of the evolutionary history and fate of a specific 
white dwarf multi-planet system.

\section*{Acknowledgements}

We thank the referee for their useful comments, which have improved the manuscript.
DV gratefully acknowledges the support of the STFC via an Ernest Rutherford Fellowship (grant ST/P003850/1), and has received, along with BTG, funding from the European Research Council under the European Union's Seventh Framework Programme (FP/2007-2013)/ERC Grant Agreement n. 320964 (WDTracer).

\label{lastpage}

\begin{thebibliography}{99}

\bibitem[Aannestad et al.(1993)]{aanetal1993} Aannestad, P.~A., Kenyon, S.~J., Hammond, G.~L., \& Sion, E.~M.\ 1993, AJ, 105, 1033 

\bibitem[Adams et al.(2013)]{adaetal2013} Adams, F.~C., Anderson, K.~R., \& Bloch, A.~M.\ 2013, MNRAS, 432, 438 

\bibitem[Althaus et al.(2010)]{altetal2010} Althaus, L.~G., C{\'o}rsico, A.~H., Isern, J., \& Garc{\'{\i}}a-Berro, E.\ 2010, ARA\&A, 18, 471 

\bibitem[Antoniadou \& Veras(2016)]{antver2016} Antoniadou, K.~I., \& Veras, D.\ 2016, MNRAS, 463, 4108 
  
\bibitem[Barnes \& Greenberg(2006)]{bargre2006} Barnes, R., \& Greenberg, R.\ 2006, ApJL, 647, L163 

\bibitem[Barnes \& Greenberg(2007)]{bargre2007} Barnes, R., \& Greenberg, R.\ 2007, ApJL, 665, L67 

\bibitem[Bear \& Soker(2013)]{beasok2013} Bear, E., \& Soker, N.\ 2013, New Astronomy, 19, 56 

\bibitem[Bonsor et al.(2011)]{bonetal2011} Bonsor, A., Mustill, A.~J., \& Wyatt, M.~C.\ 2011, MNRAS, 414, 930 

\bibitem[Bonsor \& Veras(2015)]{bonver2015} Bonsor, A., \& Veras, D.\ 2015, MNRAS, 454, 53 

\bibitem[Brown et al.(2017)]{broetal2017} Brown, J.~C., Veras, D., \& G{\"a}nsicke, B.~T.\ 2017, MNRAS, 468, 1575 

\bibitem[Butler et al.(1999)]{butetal1999} Butler, R.~P., Marcy, G.~W., Fischer, D.~A., Brown, T.~M., Contos, A.~R., Korzennik, S.~G., Nisenson, P., Noyes, R.~W.\ 1999, ApJ, 526, 916 

\bibitem[Chambers et al.(1996)]{chaetal1996} Chambers, J.~E., Wetherill, G.~W., \& Boss, A.~P.\ 1996, Icarus, 119, 261 

\bibitem[Chambers(1999)]{chambers1999} Chambers, J.~E.\ 1999, MNRAS, 304, 793 

\bibitem[Debes \& Sigurdsson(2002)]{debsig2002} Debes, J.~H., \& Sigurdsson, S.\ 2002, ApJ, 572, 556 

\bibitem[Debes et al.(2012)]{debetal2012} Debes, J.~H., Walsh, K.~J., \& Stark, C.\ 2012, ApJ, 747, 148 

\bibitem[Deck et al.(2013)]{decetal2013} Deck, K.~M., Payne, M., \& Holman, M.~J.\ 2013, ApJ, 774, 129 

\bibitem[Dennihy et al.(2018)]{denetal2018} Dennihy, E., Clemens, J.~C., Dunlap, B.~H., Fanale, S.~M., Fuchs, J.~T., Hermes, J.~J.\ 2018, ApJ, 854, 40 

\bibitem[Dong et al.(2010)]{donetal2010} Dong, R., Wang, Y., Lin, D.~N.~C., \& Liu, X.-W.\ 2010, ApJ, 715, 1036 

\bibitem[Donnison(2006)]{donnison2006} Donnison, J.~R.\ 2006, MNRAS, 369, 1267 

\bibitem[Donnison(2009)]{donnison2009} Donnison, J.~R.\ 2009, Planetary and Space Science, 57, 771 

\bibitem[Donnison(2011)]{donnison2011} Donnison, J.~R.\ 2011, MNRAS, 415, 470 

\bibitem[Dufour et al.(2007)]{dufetal2007} Dufour, P., Bergeron, P., Liebert, J., et al.\ 2007, ApJ, 663, 1291

\bibitem[Fang \& Margot(2013)]{fanmar2013} Fang, J., \& Margot, J.-L.\ 2013, ApJ, 767, 115

\bibitem[Farihi et al.(2010)]{faretal2010} Farihi, J., Barstow, M.~A., Redfield, S., Dufour, P., \& Hambly, N.~C.\ 2010, MNRAS, 404, 2123 

\bibitem[Farihi(2016)]{farihi2016} Farihi, J.\ 2016, New Astronomy Reviews, 71, 9 

\bibitem[Frewen \& Hansen(2014)]{frehan2014} Frewen, S.~F.~N., \& Hansen, B.~M.~S.\ 2014, MNRAS, 439, 2442 

\bibitem[Friedrich et al.(2004)]{frietal2004} Friedrich, S., Jordan, S., \& Koester, D.\ 2004, A\&A, 424, 665 

\bibitem[G{\"a}nsicke et al.(2006)]{gaeetal2006} G{\"a}nsicke, B.~T., Marsh, T.~R., Southworth, J., \& Rebassa-Mansergas, A.\ 2006, Science, 314, 1908 

\bibitem[G{\"a}nsicke et al.(2012)]{gaeetal2012} G{\"a}nsicke, B.~T., Koester, D., Farihi, J., et al.\ 2012, MNRAS, 424, 333 

\bibitem[Gentile Fusillo et al.(2015)]{genetal2015} Gentile Fusillo, N.~P., G{\"a}nsicke, B.~T., \& Greiss, S.\ 2015, MNRAS, 448, 2260

\bibitem[Georgakarakos(2008)]{georgakarakos2008} Georgakarakos, N.\ 2008, Celestial Mechanics and Dynamical Astronomy, 100, 151

\bibitem[Georgakarakos(2013)]{georgakarakos2013} Georgakarakos, N.\ 2013, New Astronomy, 23, 41 

\bibitem[Graham et al.(1990)]{graetal1990} Graham, J.~R., Matthews, K., Neugebauer, G., \& Soifer, B.~T.\ 1990, ApJ, 357, 216 

\bibitem[Grishin et al.(2017)]{grietal2017} Grishin, E., Perets, H.~B., Zenati, Y., \& Michaely, E.\ 2017, MNRAS, 466, 276 

\bibitem[Hadden \& Lithwick(2018)]{hadlit2018} Hadden, S., \& Lithwick, Y.\ 2018, arXiv:1803.08510

\bibitem[Hadjidemetriou(1963)]{hadjidemetriou1963} Hadjidemetriou, J.~D.\ 1963, Icarus, 2, 440 

\bibitem[Hamers \& Portegies Zwart(2016)]{hampor2016} Hamers, A.~S., \& Portegies Zwart, S.~F.\ 2016, MNRAS, 462, L84 

\bibitem[Harrison et al.(2018)]{haretal2018} Harrison, J.~H.~D., Bonsor, A., \& Madhusudhan, N.\ 2018, MNRAS, 479, 3814.

\bibitem[Hogg et al.(2018)]{hogetal2018} Hogg, M.~A., Wynn, G.~A., \& Nixon, C.\ 2018, MNRAS, 479, 4486 

\bibitem[Hollands et al.(2017)]{holetal2017} Hollands, M.~A., Koester, D., Alekseev, V., Herbert, E.~L., \& G{\"a}nsicke, B.~T.\ 2017, MNRAS, 467, 4970 

\bibitem[Hollands et al.(2018)]{holetal2018} Hollands, M.~A., G{\"a}nsicke, B.~T., \& Koester, D.\ 2018, MNRAS, 477, 93.

\bibitem[Hurley et al.(2000)]{huretal2000} Hurley, J.~R., Pols, O.~R., \& Tout, C.~A.\ 2000, MNRAS, 315, 543 

\bibitem[Hwang et al.(2017)]{hwaetal2017} Hwang, J.~A., Steffen, J.~H., Lombardi, J.~C., Jr., \& Rasio, F.~A.\ 2017, MNRAS, 470, 4145 

\bibitem[Jura(2003)]{jura2003} Jura, M.\ 2003, ApJL, 584, L91 

\bibitem[Jura(2006)]{jura2006} Jura, M.\ 2006, ApJ, 653, 613 

\bibitem[Jura \& Young(2014)]{juryou2014} Jura, M., \& Young, E.~D.\ 2014, Annual Review of Earth and Planetary Sciences, 42, 45 

\bibitem[Katz(2018)]{katz2018} Katz, J.~I.\ 2018, MNRAS, 478, L95 

\bibitem[Kenyon \& Bromley(2017a)]{kenbro2017a} Kenyon, S.~J., \& Bromley, B.~C.\ 2017a, ApJ, 844, 116 

\bibitem[Kenyon \& Bromley(2017b)]{kenbro2017b} Kenyon, S.~J., \& Bromley, B.~C.\ 2017b, ApJ, 850, 50 

\bibitem[Kepler et al.(2015)]{kepetal2015} Kepler, S.~O., Pelisoli, I., Koester, D., et al.\ 2015, MNRAS, 446, 4078 

\bibitem[Kepler et al.(2016)]{kepetal2016} Kepler, S.~O., Pelisoli, I., Koester, D., et al.\ 2016, MNRAS, 455, 3413

\bibitem[Kilic \& Redfield(2007)]{kilred2007} Kilic, M., \& Redfield, S.\ 2007, ApJ, 660, 641 

\bibitem[Klein et al.(2010)]{kleetal2010} Klein, B., Jura, M., Koester, D., Zuckerman, B., \& Melis, C.\ 2010, ApJ, 709, 950 

\bibitem[Kleinman et al.(2013)]{kleetal2013} Kleinman, S.~J., Kepler, S.~O., Koester, D., et al.\ 2013, ApJS, 204, 5

\bibitem[Koester(2013)]{koester2013} Koester, D.\ 2013, Planets, Stars and Stellar Systems.~Volume 4: Stellar Structure and Evolution, 4, 559 

\bibitem[Koester et al.(2014)]{koeetal2014} Koester, D., G{\"a}nsicke, B.~T., \& Farihi, J.\ 2014, A\&A, 566, A34 

\bibitem[Kostov et al.(2016)]{kosetal2016} Kostov, V.~B., Moore, K., Tamayo, D., Jayawardhana, R., \& Rinehart, S.~A.\ 2016, ApJ, 832, 183 

\bibitem[Kozai(1962)]{kozai1962} Kozai, Y.\ 1962, AJ, 67, 591 

\bibitem[Kratter \& Perets(2012)]{kraper2012} Kratter, K.~M., \& Perets, H.~B.\ 2012, ApJ, 753, 91 

\bibitem[Kratter \& Shannon(2014)]{krasha2014} Kratter, K.~M., \& Shannon, A.\ 2014, MNRAS, 437, 3727 

\bibitem[Lidov(1962)]{lidov1962} Lidov, M.~L.\ 1962, Planetary and Space Science, 9, 719 

\bibitem[Luhman et al.(2011)]{luhetal2011} Luhman, K.~L., Burgasser, A.~J., \& Bochanski, J.~J.\ 2011, ApJL, 730, L9 

\bibitem[Manser et al.(2016)]{manetal2016} Manser, C.~J., G{\"a}nsicke, B.~T., Marsh, T.~R., et al.\ 2016, MNRAS, 455, 4467 

\bibitem[Marchal \& Bozis(1982)]{marboz1982} Marchal, C., \& Bozis, G.\ 1982, Celestial Mechanics, 26, 311 

\bibitem[Mardling(2008)]{mardling2008} Mardling, R.~A.\ 2008, Dynamical Evolution of Dense Stellar Systems, 246, 199 

\bibitem[Marzari(2014)]{marzari2014} Marzari, F.\ 2014, MNRAS, 442, 1110 

\bibitem[Metzger et al.(2012)]{metetal2012} Metzger, B.~D., Rafikov, R.~R., \& Bochkarev, K.~V.\ 2012, MNRAS, 423, 505 

\bibitem[Miranda \& Rafikov(2018)]{mirraf2018} Miranda, R., \& Rafikov, R.~R.\ 2018, ApJ, 857, 135.

\bibitem[Mustill et al.(2013)]{musetal2013} Mustill, A.~J., Marshall, J.~P., Villaver, E., et al.\ 2013, MNRAS, 436, 2515 

\bibitem[Mustill et al.(2014)]{musetal2014} Mustill, A.~J., Veras, D., \& Villaver, E.\ 2014, MNRAS, 437, 1404 

\bibitem[Mustill et al.(2018)]{musetal2018} Mustill, A.~J., Villaver, E., Veras, D.,  G{\"a}nsicke, B.~T., Bonsor, A. \ 2018, MNRAS, 476, 3939.

\bibitem[Naoz(2016)]{naoz2016} Naoz, S.\ 2016, ARA\&A, 54, 441 

\bibitem[Omarov(1962)]{omarov1962} Omarov, T.~B. 1962, Izv. Astrofiz. Inst. Acad. Nauk. KazSSR, 14, 66

\bibitem[Payne et al.(2016)]{payetal2016} Payne, M.~J., Veras, D., Holman, M.~J., G\"{a}nsicke, B.~T.\ 2016, MNRAS, 457, 217 

\bibitem[Payne et al.(2017)]{payetal2017} Payne, M.~J., Veras, D., G{\"a}nsicke, B.~T., \& Holman, M.~J.\ 2017, MNRAS, 464, 2557 

\bibitem[Perryman et al.(2014)]{peretal2014} Perryman, M., Hartman, J., Bakos, G. {\'A}., Lindegren, L.\ 2014, ApJ, 797, 14.

\bibitem[Petit et al.(2018)]{petetal2018} Petit, A.~C., Laskar, J., \& Bou{\'e}, G.\ 2018, arXiv:1806.08869 

\bibitem[Petrovich(2015)]{petrovich2015} Petrovich, C.\ 2015, ApJ, 808, 120 

\bibitem[Petrovich \& Mu{\~n}oz(2017)]{petmun2017} Petrovich, C., \& Mu{\~n}oz, D.~J.\ 2017, ApJ, 834, 116 

\bibitem[Pichierri et al.(2017)]{picetal2017} Pichierri, G., Morbidelli, A., \& Lai, D.\ 2017, A\&A, 605, A23 
  
\bibitem[Portegies Zwart(2013)]{portegieszwart2013} Portegies Zwart, S.\ 2013, MNRAS, 429, L45 

\bibitem[Pu \& Wu(2015)]{puwu2015} Pu, B., \& Wu, Y.\ 2015, ApJ, 807, 44 

\bibitem[Quillen(2011)]{quillen2011} Quillen, A.~C.\ 2011, MNRAS, 418, 1043 

\bibitem[Rafikov(2011a)]{rafikov2011a} Rafikov, R.~R.\ 2011a, MNRAS, 416, L55 

\bibitem[Rafikov(2011b)]{rafikov2011b} Rafikov, R.~R.\ 2011b, ApJL, 732, L3 

\bibitem[Ramos et al.(2015)]{rametal2015} Ramos, X.~S., Correa-Otto, J.~A., \& Beaug{\'e}, C.\ 2015, Celestial Mechanics and Dynamical Astronomy, 123, 453 

\bibitem[Raymond et al.(2009)]{rayetal2009} Raymond, S.~N., Barnes, R., Veras, D., et al.\ 2009, ApJL, 696, L98 

\bibitem[Schatzman(1958)]{schatzman1958} Schatzman, E.~L.\ 1958, Amsterdam, North-Holland Pub.~Co.; New York, Interscience Publishers, 1958  

\bibitem[Schleicher \& Dreizler(2014)]{schdre2014} Schleicher, D.~R.~G., \& Dreizler, S.\ 2014, A\&A, 563, A61 

\bibitem[Schr{\"o}der \& Cuntz(2005)]{schcun2005} Schr{\"o}der, K.-P., \& Cuntz, M.\ 2005, ApJ, 630, L73.

\bibitem[Shannon et al.(2014)]{shaetal2014} Shannon, A., Clarke, C., \& Wyatt, M.\ 2014, MNRAS, 442, 142.

\bibitem[Sigurdsson et al.(2003)]{sigetal2003} Sigurdsson, S., Richer, H.~B., Hansen, B.~M., Stairs, I.~H., \& Thorsett, S.~E.\ 2003, Science, 301, 193 

\bibitem[Smallwood et al.(2018)]{smaetal2018} Smallwood, J.~L., Martin, R.~G., Livio, M., \& Lubow, S.~H.\ 2018, MNRAS, 480, 57 

\bibitem[Stephan et al.(2017)]{steetal2017} Stephan, A.~P., Naoz, S., \& Zuckerman, B.\ 2017, ApJL, 844, L16 

\bibitem[Stephan et al.(2018)]{steetal2018} Stephan, A.~P., Naoz, S., \& Gaudi, B.~S.\ 2018, Submitted to AAS Journals, arXiv:1806.04145 

\bibitem[Tremblay et al.(2016)]{treetal2016} Tremblay, P.-E., Cummings, J., Kalirai, J.~S., G\"{a}nsicke, B.~T., Gentile-Fusillo, N., Raddi, R. \ 2016, MNRAS, 461, 2100 

\bibitem[van Lieshout et al.(2018)]{vanetal2018} van Lieshout, R., Kral, Q., Charnoz, S., et al.\ 2018, MNRAS, 480, 2784.

\bibitem[Vanderburg et al.(2015)]{vanetal2015} Vanderburg, A., Johnson, J.~A., Rappaport, S., et al.\ 2015, Nature, 526, 546 

\bibitem[Vassiliadis \& Wood(1993)]{vaswoo1993} Vassiliadis, E., \& Wood, P.~R.\ 1993, ApJ, 413, 641

\bibitem[Veras \& Armitage(2004)]{verarm2004} Veras, D., \& Armitage, P.~J.\ 2004, Icarus, 172, 349 

\bibitem[Veras(2007)]{veras2007} Veras, D.\ 2007, Celestial Mechanics and Dynamical Astronomy, 99, 197 
  
\bibitem[Veras et al.(2011)]{veretal2011} Veras, D., Wyatt, M.~C., Mustill, A.~J., Bonsor, A., \& Eldridge, J.~J.\ 2011, MNRAS, 417, 2104 

\bibitem[Veras \& Tout(2012)]{vertou2012} Veras, D., \& Tout, C.~A.\ 2012, MNRAS, 422, 1648 

\bibitem[Veras \& Wyatt(2012)]{verwya2012} Veras, D., \& Wyatt, M.~C.\ 2012, MNRAS, 421, 2969

\bibitem[Veras \& Evans(2013)]{vereva2013} Veras, D., \& Evans, N.~W.\ 2013, MNRAS, 430, 403.

\bibitem[Veras et al.(2013a)]{veretal2013a} Veras, D., Mustill, A.~J., Bonsor, A., \& Wyatt, M.~C.\ 2013a, MNRAS, 431, 1686 

\bibitem[Veras et al.(2013b)]{veretal2013b} Veras, D., Hadjidemetriou, J.~D., \& Tout, C.~A.\ 2013b, MNRAS, 435, 2416 

\bibitem[Veras \& Mustill(2013)]{vermus2013} Veras, D., \& Mustill, A.~J.\ 2013, MNRAS, 434, L11 

\bibitem[Veras et al.(2014a)]{veretal2014a} Veras, D., Leinhardt, Z.~M., Bonsor, A., G\"{a}nsicke, B.~T.\ 2014a, MNRAS, 445, 2244
 
\bibitem[Veras et al.(2014b)]{veretal2014b} Veras, D., Jacobson, S.~A., G\"{a}nsicke, B.~T.\ 2014b, MNRAS, 445, 2794 

\bibitem[Veras et al.(2014c)]{veretal2014c} Veras, D., Shannon, A., \& G{\"a}nsicke, B.~T.\ 2014c, MNRAS, 445, 4175.

\bibitem[Veras et al.(2014d)]{veretal2014d} Veras, D., Evans, N.~W., Wyatt, M.~C., et al.\ 2014d, MNRAS, 437, 1127.

\bibitem[Veras \& G\"{a}nsicke(2015)]{vergae2015} Veras, D., G\"{a}nsicke, B.~T.\ 2015, MNRAS, 447, 1049 

\bibitem[Veras et al.(2015a)]{veretal2015a} Veras, D., Leinhardt, Z.~M., Eggl, S., G{\"a}nsicke, B.~T.\ 2015a, MNRAS, 451, 3453

\bibitem[Veras et al.(2015b)]{veretal2015b} Veras, D., Eggl, S., G{\"a}nsicke, B.~T.\ 2015b, MNRAS, 451, 2814 

\bibitem[Veras(2016a)]{veras2016a} Veras, D.\ 2016a, Royal Society Open Science, 3, 150571 

\bibitem[Veras(2016b)]{veras2016b} Veras, D.\ 2016b, MNRAS, 463, 2958 

\bibitem[Veras et al.(2016)]{veretal2016} Veras, D., Mustill, A.~J., G{\"a}nsicke, B.~T., et al.\ 2016, MNRAS, 458, 3942 

\bibitem[Veras et al.(2017a)]{veretal2017a} Veras, D., Georgakarakos, N., Dobbs-Dixon, I., \& G{\"a}nsicke, B.~T.\ 2017a, MNRAS, 465, 2053 

\bibitem[Veras et al.(2017b)]{veretal2017b} Veras, D., Mustill, A.~J., \& G{\"a}nsicke, B.~T.\ 2017b, MNRAS, 465, 1499 
  
\bibitem[Veras et al.(2018)]{veretal2018} Veras, D., Xu, S., \& Rebassa-Mansergas, A.\ 2018, MNRAS, 473, 2871 

\bibitem[V{\"o}lschow et al.(2014)]{voletal2014} V{\"o}lschow, M., Banerjee, R., \& Hessman, F.~V.\ 2014, A\&A, 562, A19 

\bibitem[Voyatzis et al.(2013)]{voyetal2013} Voyatzis, G., Hadjidemetriou, J.~D., Veras, D., \& Varvoglis, H.\ 2013, MNRAS, 430, 3383 

\bibitem[Wilson et al.(2015)]{wiletal2015} Wilson, D.~J., G{\"a}nsicke, B.~T., Koester, D., Toloza, O., Pala, A.~F., Breedt, E., Parsons, S.~G. \ 2015, MNRAS, 451, 3237 

\bibitem[Wilson et al.(2016)]{wiletal2016} Wilson, D.~J., G{\"a}nsicke, B.~T., Farihi, J., \& Koester, D.\ 2016, MNRAS, 459, 3282 

\bibitem[Wisdom(1980)]{wisdom1980} Wisdom, J.\ 1980, AJ, 85, 1122 

\bibitem[Wyatt et al.(2014)]{wyaetal2014} Wyatt, M.~C., Farihi, J., Pringle, J.~E., \& Bonsor, A.\ 2014, MNRAS, 439, 3371 

\bibitem[Xu \& Jura(2014)]{xujur2014} Xu, S., \& Jura, M.\ 2014, ApJL, 792, L39 

\bibitem[Xu et al.(2017)]{xuetal2017} Xu, S., Zuckerman, B., Dufour, P., Young, E.~D., Klein, B., Jura, M.\ 2017, ApJL, 836, L7 

\bibitem[Zuckerman et al.(2003)]{zucetal2003} Zuckerman, B., Koester, D., Reid, I.~N., H\"{u}nsch, M.\ 2003, ApJ, 596, 477 

\bibitem[Zuckerman et al.(2010)]{zucetal2010} Zuckerman, B., Melis, C., Klein, B., Koester, D., \& Jura, M.\ 2010, ApJ, 722, 725 

\end{thebibliography}
\end{document}